# Pure bound field theory and structure of atomic energy levels

A.L. Kholmetskii

*Department of Physics, Belarus State University, 4 Nezavisimosti Avenue, 220030 Minsk, Belarus & Okan University, Istanbul, Turkey*

O.V. Missevitch

*Institute for Nuclear Problems, Belarus State University, 11 Bobruiskaya Str., 220030 Minsk, Belarus*

T. Yarman

*Department of Engineering, Okan University, Akfirat, Istanbul, Turkey & Savronik, Eskisehir, Turkey*

In this paper we continue the analysis of quantum two-particle bound systems we have started in our earlier contribution (Kholmetskii, A.L., Missevitch, O.V. and Yarman, T. *Phys. Scr*., **82** (2010), 045301), where we re-postulated the Dirac equation for the bound electron in an external EM field based on the requirement of total momentum conservation, when its EM radiation is prohibited. It has been shown that the modified expression for the energy levels of hydrogenic atoms within such a pure bound field theory (PBFT) provides the same gross and fine structure of energy levels that had been predicted by the standard theory. Now we apply the PBFT to the analysis of hyperfine interactions and show the appearance of some important corrections to the energy levels (the 1*S*-2*S* interval and hyperfine spin-spin splitting in positronium, 1*S* and 2*S*-2*P* Lamb shift in hydrogen), which remedies considerably the discrepancy between theoretical predictions and experimental results. In particular, the corrected 1*S*-2*S* interval and the spin-spin splitting in positronium practically eliminate the available up to date deviation between theoretical and experimental data. The re-estimated classic 2*S*-2*P* Lamb shift as well as ground state Lamb shift in the hydrogen atom lead to the proton charge radius $r_p$=0.841(6) fm (from 2*S*-2*P* Lamb shift), and $r_p$=0.844(24) fm (from 1S Lamb shift), which perfectly agress with the latest estimation of proton size via the measurement of 2*S*-2*P* Lamb shift in muonic hydrogen, i.e. $r_p$=0.84184(67) fm. We also emphasize the universal character of PBFT, which is applicable to heavy atoms, too, and analyze 2*S*-2*P* interval in Li-like uranium. We show that the corrections we introduced provide a better correspondence between the calculated and experimental data than that furnished by the standard approach. The results obtained support our principal idea of the enhancement of the bound EM field in the absence of EM radiation for quantum bound systems.

## 1. Introduction

The development of quantum electrodynamics (QED) in the second half of the past century and the unexampled success of this theory in physics of light hydrogenic atoms (where the deviation be-tween theoretical predictions and experimental results gradually reaches relative values of $10^{-12}$, and even $10^{-13}$), completely convinced physicists about the outstanding correctness of QED with regards to the description of quantum phenomena. At the same time, modern progress both in the theory and experiment allowed recently the disclosure of a number of unexplained deviations between theoretical predictions and experimental data in physics of light hydrogenic atoms, where the value of such deviation ($\Delta$) substantially exceeds the corresponding uncertainty $\sigma$ (both theoretical and experimental). Currently the revealed discrepancies between QED calculations and experimental results are not the subject of wide scientific discussions, and thus it is worth to remind them below:

- 1*S*-2*S* interval in positronium ($\Delta/\sigma \approx 3.0$) [1, 2];
- hyperfine interval in positronium ($\Delta/\sigma \approx 2.5$) [1];
- proton charge radius $r_p$ derived from the classic 2*S*-2*P* Lamb shift and the ground state Lamb shift in hydrogen systematically exceed the value of $r_p$ obtained from particle physics ($\Delta/\sigma$ varies from 3 to 5 according to different estimations [3, 4]).

One can add the result of very recent experiment [5], where the proton charge radius being estimated via the measurements of 2*S*-2*P* Lamb shift in muonic hydrogen ($r_p$=0.84184(67) fm), occurred to be less by 5.0 standard deviations than the modern CODATA value $r_p$=0.8768(69) fm [6]. Moreover, for muonic hydrogen the nuclear size effect contributes significantly (about 2 %) to the 2*S*-2*P* Lamb shift and thus this new value of $r_p$ can pretend to be the most precise amongst all available results. If so, its deviation from the value of $r_p$ extracted via the 2*S*-2*P* Lamb shift and 1*S* Lamb shift in the hydrogen occurs to be drastic.

Thus, these facts may indicate on a possible presence of some still missed elements in the description of quantum two-particle bound systems.

In this respect we mention our recent paper [7], where we pointed out the known fact that quantum bound charges do not radiate at stationary energy states and thus their EM field consists of the bound (non-radiative) component only. This effect does not have a classical analogy, where, as known, an orbiting charge must inevitably radiate, and both bound and radiative EM field components equally participate in securing the total momentum conservation law. Hence the question emerges, which appears to be has not been asked before ref. [7]: How does Nature restore the energy-momentum conservation law for quantum *bound* systems of charges, where their EM radiation is prohibited?

As the general approach to the analysis of this problem, one can consider QED equations for Dirac field and EM field (the non-homogeneous wave equation for the four-potential [8]) and, using the formalism developed in ref. [9], to modify this wave equation for bound four-potential, canceling by such a way EM radiation. However, in the paper [7] we applied a much more simple but more illustrative way based on the analysis of one-body problem in the classical limit (still ignoring spin effects), via further taking into account the Bohr's correspondence principle.

From the general viewpoint, it is clear that a possible way to implement the energy-momentum conservation law for an isolated system of charges with the prohibited EM radiation is to modify in an appropriate way their bound EM field (potentials) and/or the relationships between the fields (potentials) and their energy-momentum. Exploring this problem and considering the electron in an external EM field, we first have to distinguish its states with the total energy $E \geq mc^2$ (free electron, where $m$ is its rest mass and $c$ the light velocity in vacuum), and $E < mc^2$ (bound electron).

At $E \geq mc^2$ the electron's energy spectrum is continuous, and its radiative EM field is not prohibited. In contrast, at $E < mc^2$ the electron is characterized by the discrete energy spectrum, and it cannot emit EM radiation in a stationary energy state. Hence any possible modifications of the Dirac equation (DE) aimed to take into account the momentum conservation constraint for non-radiative nature of EM field of bound electron, should be made via the introduction of some step-wise func-tion of the difference ($E$-$mc^2$), which yields the common DE at $E \geq mc^2$. At the same time, it is reasonable to assume that for a bound electron ($E<mc^2$), the implementation of the total momentum conservation law in the absence of radiating EM field component would require the appropriate modifications of expressions for momentum of EM field and/or interaction EM energy. Thus the general form of DE for the electron in an external EM field, which reflects such modifications of bound EM field at $E<mc^2$, reads [7]:

$$i\hbar \frac{\partial}{\partial t}\psi = \left[ c\,\boldsymbol{\alpha}\left(\hat{\boldsymbol{P}}_b - \frac{e}{c}\boldsymbol{A}_b\right) + \beta m B_n c^2 + \Gamma_n e\varphi \right]\psi ,$$

where $\psi = \begin{pmatrix} \xi \\ \chi \end{pmatrix}$ is the wave function, $\boldsymbol{\alpha} = \begin{pmatrix} 0 & \sigma \\ \sigma & 0 \end{pmatrix}$, $\beta = \begin{pmatrix} 1 & 0 \\ 0 & -1 \end{pmatrix}$, $\hat{\boldsymbol{P}}_b = -i\hbar\nabla$ is the operator of canonical momentum, $e$ is the electron's charge, $\boldsymbol{A}_b$ is the vector potential, $\boldsymbol{\sigma}$ is the Pauli matrix, $\varphi$ is the scalar potential and the step-wise functions $B_n$, $\Gamma_n$ are defined by the relationships

$$B_n\left(E - mc^2\right) = \begin{matrix} 1\,(at\ E \geq mc^2) \\ b_n(at\ E < mc^2) \end{matrix}, \quad \Gamma_n\left(E - mc^2\right) = \begin{matrix} 1\,(at\ E \geq mc^2) \\ \gamma_n(at\ E < mc^2) \end{matrix},$$

where $b_n$, $\gamma_n$ are some coefficients, we introduced. These coefficients are aimed to reflect the non-radiative nature of EM field of bound electron within the total momentum conservation constraint and thus, in the non-relativistic limit both $b_n$ and $\gamma_n$ are put to be equal to unity. Hence their introduction in DE does not affect the gross structure of the electron's energy levels, characterized by the principal quantum number $n$. These coefficients, being constant for any fixed energy level, provide the Lorentz-invariance of the modified DE, when the non-relativistic limit is no longer assumed [7]. From the physical viewpoint, the presence of step-wise functions $B_n$ and $\Gamma_n$ in the Dirac equation reflects a discontinuity of the properties of the electron, which can emit EM radiation at the total energy $E \geq mc^2$, but loses the ability to radiate at $E<mc^2$.

In order to find the explicit form of the coefficients $b_n$ and $\gamma_n$ for bound electron, we considered in ref. [7] a model classical one-body problem, where EM radiation of the electron orbiting around a heavy nucleus, is prohibited, and further determined the way of modifying the bound EM field of this system, which guarantees the implementation of the total momentum conservation law. By such a way we found the classical limits for the coefficient $b_n$ and $\gamma_n$, which occurred sufficient to find $b_n$ and $\gamma_n$ themselves [7].

Deriving further the modified Dirac-Coulomb equation for the quantum one-body problem on the basis of modified DE, we arrived at the same gross, also fine structure of energy levels, as those furnished by the conventional approach, for hydrogenlike atoms, but obtained a small difference in the value of average radius of electron's orbit [7]. A relative change of the radius of the electron orbits for hydrogen-like atoms has the order of magnitude $(Z\alpha)^2$ (where $Z$ is the atomic number and $\alpha$ the fine structure constant), and is substantially smaller than the present experimental uncertainly in the measurement of atomic form-factors [1].

In the present paper we extend the approach of ref. [7] based on modified Dirac equation and named hereinafter as Pure Bound Field Theory (PBFT) to quantum two-body problems (section 2) and further analyze the hyperfine contributions to the atomic energy levels (section 3), which, in addition to the modified solution of quantum mechanical equations for the two-body problems, yields the replacement $U \rightarrow \gamma_n U$ in the input of QED expressions for radiative corrections [7], without altering the core structure of this theory. We show that the corrections of PBFT to fine structure (having the order of magnitude of hyperfine interactions) are significant only for 1$S$ states of hydro-genic atoms (sub-section 3.1). The correction brought by the PBFT to the hyperfine spin-spin interaction is completely negligible for the hydrogen-like atoms, except positronium, where such a correction becomes significant (sub-section 3.2) to eliminate the available disagreement between calculated and experimental data. In subsection 3.3 we derive the PBFT corrections to 1*S* and 2*S* Lamb shift, which substantially exceed the uncertainly (both theoretical and experimental) in their determination for light hydrogenlike atoms. In section 4 we show that the corrections brought by the PBFT to common results practically eliminate totally the available up to date discrepancy between theoretical and experimental data for the 1*S*-2*S* interval in positronium, spin-spin splitting in positronium, classic Lamb shift and ground state Lamb shift in hydrogen. In particular, we derive the proton charge radius $r_p$=0.837(8) fm (from 2*S*-2*P* Lamb shift), $r_p$=0.840(24) fm (from 1*S* Lamb shift), which completely agrees with the latest experimental result [5]. We also emphasize the universal character of our approach, which is thus applicable to heavy atoms, too, and analyze the 2*S*-2*P* interval in Li-like uranium within the framework of PBFT (section 5), where the corrections we introduce provide a better correspondence between the calculated and experimental data than that furnished by the common approach. Finally, we conclude in section 6.

## 2. Classical and quantum two-body problem

In this section we consider first the hydrogenlike atom in the semi-classical treatment, suggested in [7], when the motion of electron with the charge *e* and mass *m* around the nucleus with the charge *Ze* and mass *M* is described in the classical way, but the EM radiation of this system is prohibited, like in the quantum case. For the one-body problem ($M\to\infty$) we obtained the motional equation in the form:

$$\frac{d}{dt}\gamma\left(m+\frac{U_1}{c^2}\right)\boldsymbol{v}=-\gamma\frac{Ze^2\boldsymbol{r}}{r^3},$$

which occurs sufficient to determine the classical limits of coefficients $b_n$ ($b=1+U/mc^2$) and $\gamma_n$ ($\gamma=\left(1-v^2/c^2\right)^{-1/2}$) in the modified DE. Here $U_1=\frac{1}{4\pi}\int\limits_V \boldsymbol{E}_e\cdot\boldsymbol{E}_N dV=-\frac{Ze^2}{r}$ is the Coulomb interaction EM energy for the one-body problem; $\boldsymbol{E}_e$ and $\boldsymbol{E}_N$ stand for the electric fields of electron and nucleus, correspondingly, and the integration is carried out over the entire three-space *V*.

Analogously, when the mass of the nucleus *M* takes the finite value (two-body problem), we get the motional equation

$$\frac{d}{dt}\gamma_m\left(m+\gamma_M U/c^2\right)\boldsymbol{v}_m=-\frac{d}{dt}\gamma_M\left(M+\gamma_m U/c^2\right)\boldsymbol{v}_M, \qquad (1)$$

where

$$U=\frac{1}{4\pi}\int\limits_V\left(\boldsymbol{E}_m\cdot\boldsymbol{E}_M+\boldsymbol{B}_m\cdot\boldsymbol{B}_M\right)dV \qquad (2)$$

is the EM interaction energy.

Eq. (1) allows defining the *effective canonical* momenta for the electron and nucleus

$$\boldsymbol{p}_{bm}=\gamma_m\left(m+\gamma_M U/c^2\right)\boldsymbol{v}_m=\gamma_m m_b\boldsymbol{v}_m=b_m\boldsymbol{p}_m,\ \ \boldsymbol{p}_{bM}=\gamma_M\left(M+\gamma_m U/c^2\right)\boldsymbol{v}_M=\gamma_M M_b\boldsymbol{v}_M=b_M\boldsymbol{p}_M, \qquad (2a\text{-}b)$$

where we introduce the *effective* rest masses

$$m_b = m\left(1+\gamma_M U/mc^2\right) = mb_m\,,\ M_b = M\left(1+\gamma_m U/Mc^2\right) = Mb_M \qquad \text{(3a-b)}$$

of both particles and designated $\boldsymbol{p}_m$, $\boldsymbol{p}_M$ the mechanical momenta of electron and nucleus, correspondingly. Herein, like in ref. [7], we supply the quantities of PBFT by the subscript "$b$", and introduce the factors

$$b_m = \left(1+\frac{\gamma_M U}{mc^2}\right),\ b_M = \left(1+\frac{\gamma_m U}{Mc^2}\right),\ \gamma_m = \left(1-v_m^{\ 2}/c^2\right)^{-1/2},\ \gamma_M = \left(1-v_M^{\ 2}/c^2\right)^{-1/2}. \quad \text{(4a-d)}$$

Then the Hamilton function, written in the weak relativistic limit to the accuracy $c^{-2}$, becomes

$$H = \frac{p_b^2}{2mb_m} + \frac{p_b^2}{2Mb_M} + \frac{p_b^4}{8m^3 b_m^{\ 3}} + \frac{p_b^4}{8M^3 b_M^{\ 3}} - \gamma_m\gamma_M \frac{Zq^2}{r}\,,\ \text{with}\ \boldsymbol{p}_b = \boldsymbol{p}_{bm} = -\boldsymbol{p}_{bM}\,.$$

The approach consisting in going from the classical description to quantum description of two-particle bound system and taking into account the Dirac equation**,** is not directly applicable to the two-particle case, where we should address either to the Bethe-Salpeter equation, or the Breit equation without external field [10], or to their appropriate modifications. Though the Breit equation is not fully Lorentz-invariant and represents an approximate, it is the most convenient and illustrative for the analysis of PBFT corrections, resulting due to the replacements

$$m \to b_{mn} m,\ M \to b_{Mn} M,\ U \to \gamma_{mn}\gamma_{Mn} U, \qquad \text{(5a-c)}$$

where the classical limits of the introduced coefficients $b_{mn}$, $b_{Mn}$, $\gamma_{mn}$ and $\gamma_{Mn}$ are determined by the respective equations (4a-d). Besides, we define the operators of canonical momenta $\hat{p}_{bm} = \hat{p}_{bM} \equiv p_b = -i\hbar\nabla$, where we take into account that due to the total momentum conservation law, $\boldsymbol{p}_{bm}=-\boldsymbol{p}_{bM} \equiv \boldsymbol{p}_b$. Hence the Breit equation modified by the replacements (5a-c) reads:

$$\left(H_{b0} + H_{b1} + \ldots + H_{b5}\right)\xi\left(\boldsymbol{r}_m, \boldsymbol{r}_M\right) = W\xi\left(\boldsymbol{r}_m, \boldsymbol{r}_M\right), \qquad (6)$$

where $\xi\left(\boldsymbol{r}_m, \boldsymbol{r}_M\right)$ is the wave function having 16 spinor components, $\boldsymbol{r}_m, \boldsymbol{r}_M$ are the position vectors for each particle, $W$ is the energy, and the Hamiltonian components $H_{bi}$ have the form:

$$H_{b0} = -\gamma_{mn}\gamma_{Mn}\frac{Ze^2}{r} + \frac{1}{2}\left(\frac{1}{b_{mn}m} + \frac{1}{b_{Mn}M}\right)p_b^{\ 2}, \qquad \text{(6a)}$$

$$H_{b1} = -\frac{1}{8c^2}\left(\frac{1}{b_{mn}^{\ 3}m^3} + \frac{1}{b_{Mn}^{\ 3}M^3}\right)p_b^{\ 4}, \qquad \text{(6b)}$$

$$H_{b2} = -\frac{Ze^2}{2b_{mn}b_{Mn}mMc^2}\frac{1}{r}\left(p_b^{\ 2} + \frac{1}{r^2}\boldsymbol{r}\cdot\left(\boldsymbol{r}\cdot\boldsymbol{p}_b\right)\boldsymbol{p}_b\right), \qquad \text{(6c)}$$

$$H_{b3} = -\gamma_{mn}\gamma_{Mn}\frac{\boldsymbol{r}\times\boldsymbol{p}_b}{r^3}\left(\frac{Ze^2\hbar}{2b_{mn}^{\ 2}m^2c^2}\boldsymbol{s}_m + \frac{Ze^2\hbar}{2b_{Mn}^{\ 2}M^2c^2}\boldsymbol{s}_M + \frac{Ze^2\hbar}{b_{mn}b_{Mn}mMc^2}\boldsymbol{s}_m + \frac{Ze^2\hbar}{b_{mn}b_{Mn}mMc^2}\boldsymbol{s}_M\right), \qquad \text{(6d)}$$

$$H_{b4} = -\frac{iZe^2\hbar}{2c^2}\left(\frac{1}{b_{mn}{}^2 m^2} + \frac{1}{b_{Mn}{}^2 M^2}\right)\boldsymbol{p}_b \cdot \nabla \frac{1}{r}, \qquad \text{(6e)}$$

$$H_{b5} = \frac{Ze^2\hbar^2\gamma_{mn}\gamma_{Mn}}{b_{mn}b_{Mn}mM}\left(-\frac{8\pi}{3}(\boldsymbol{s}_m \cdot \boldsymbol{s}_M)\delta(\boldsymbol{r}) + \frac{1}{r^3}(\boldsymbol{s}_m \cdot \boldsymbol{s}_M - 3s_{mr}s_{Mr})\right), \qquad \text{(6f)}$$

where $s_{mr} = \frac{\boldsymbol{s}_m \cdot \boldsymbol{r}}{r}$, and $\boldsymbol{r}=\boldsymbol{r}_m-\boldsymbol{r}_M$.

Here the first operator $H_{b0}$ represents the counterpart of PBFT to the conventional Schrödinger operator; $H_{b1}$ is the relativistic expansion of $H_{b0}$; the operator $H_{b2}$ takes into account the retardation in the interaction of electron and nucleus; $H_{b3}$ describes spin-orbit interaction; $H_{b4}$ is responsible for the contact interaction, and $H_{b5}$ stands for spin-spin interaction, where we take into account that in PBFT, the interaction of two spins (currents) is $\gamma_{mn}\gamma_{Mn}$ times larger than in the common theory, due to the modified field equation for magnetic field and Lorentz force law in pure bound field electrodynamics (eqs. (13d) and (11) of ref. [7], correspondingly). The same factor $\gamma_{mn}\gamma_{Mn}$ is introduced into the term of spin-orbit interaction, because the spin-orbit term originally includes the electric field of the nucleus $\boldsymbol{E}_b$, which is related to the usual electric field $\boldsymbol{E}=Ze\boldsymbol{r}/r^3$ by the relationship $\boldsymbol{E}_b \to \gamma_{mn}\gamma_{Mn}\boldsymbol{E}$, implied by eq. (5c). (In fact, the substitution $\boldsymbol{E}_b \to \gamma_{mn}\gamma_{Mn}\boldsymbol{E}$ originates from the field equations and Lorentz force law applied to two-body problem in pure bound field electrodynamics (see eqs. (13a) and (11) of ref. [7]).

For further analysis it is convenient to reduce the Breit equation to the Schrödinger-like type by analogy with ref. [11]:

$$\left[\frac{p_b^2}{2mb_{mn}} + \frac{p_b^2}{2Mb_{mn}} - \gamma_{mn}\gamma_{Mn}\frac{Ze^2}{r} - \frac{p_b^4}{8m^3 b_{mn}{}^3 c^2} - \frac{p_b^2}{8M^3 b_{Mn}{}^3 c^2} + U_b(\boldsymbol{p}_{bm}, \boldsymbol{p}_{bM}, \boldsymbol{r})\right]\psi(\boldsymbol{r}) = W\psi(\boldsymbol{r}), \qquad \text{(7)}$$

where $W$ is the energy, and the term $U_b(\boldsymbol{p}_{bm}, \boldsymbol{p}_{bM}, \boldsymbol{r})$ is equal to

$$\begin{aligned} U_b(\boldsymbol{p}_{bm}, \boldsymbol{p}_{bM}, \boldsymbol{r}) = {} & -\frac{\pi Ze^2\hbar^2}{2c^2}\left(\frac{1}{b_{mn}{}^2 m^2} + \frac{1}{b_{Mn}{}^2 M^2}\right)\delta(\boldsymbol{r}) - \frac{Ze^2}{2b_{mn}b_{Mn}mMr}\left(\boldsymbol{p}_{bm} \cdot \boldsymbol{p}_{bM} + \frac{\boldsymbol{r}\cdot(\boldsymbol{r}\cdot\boldsymbol{p}_{bm})\boldsymbol{p}_{bM}}{r^2}\right) - \\ & \frac{Ze^2\hbar\gamma_{mn}\gamma_{Mn}}{4b_{mn}{}^2 m^2 c^2 r^3}(\boldsymbol{r}\times\boldsymbol{p}_{bm})\cdot \sigma_m + \frac{Ze^2\hbar\gamma_{mn}\gamma_{Mn}}{4b_{Mn}{}^2 M^2 c^2 r^3}(\boldsymbol{r}\times\boldsymbol{p}_{bM})\cdot \sigma_M - \frac{Ze^2\hbar\gamma_{mn}\gamma_{Mn}}{2b_{mn}b_{Mn}mMc^2 r^3}\left((\boldsymbol{r}\times\boldsymbol{p}_{bm})\cdot \sigma_M - (\boldsymbol{r}\times\boldsymbol{p}_{bM})\cdot \sigma_m\right) + \\ & \frac{Ze^2\hbar\gamma_{mn}\gamma_{Mn}}{4b_{mn}b_{Mn}mMc^2}\left[\frac{\sigma_m \cdot \sigma_M}{r^3} - 3\frac{(\sigma_m\cdot\boldsymbol{r})(\sigma_M\cdot\boldsymbol{r})}{r^3} - \frac{8\pi}{3}\sigma_m \cdot \sigma_M\delta(\boldsymbol{r})\right]. \end{aligned} \qquad \text{(8)}$$

In order to solve eq. (7), it is convenient to apply the substitution

$$\boldsymbol{r} = \boldsymbol{r}'/(b_{mn}b_{Mn}\gamma_{mn}\gamma_{Mn}), \qquad \text{(9)}$$

which will allows us to present the Hamiltonian in eq. (7) as the sum of non-relativistic Schrödinger term and perturbation. Indeed, taking into account that $p_b^2 = -\hbar^2\nabla_r^2 = -b_{mn}{}^2 b_{Mn}{}^2 \gamma_{mn}{}^2 \gamma_{Mn}{}^2 \hbar^2 \nabla_{r'}^2$, we transform eq. (7) as follows:

$$\left[\begin{array}{l}-\dfrac{\hbar^2\nabla_{r'}^2 b_{Mn}}{2m}-\dfrac{\hbar^2\nabla_{r'}^2 b_{mn}}{2M}-\dfrac{Ze^2}{r'}+\\ \dfrac{1}{b_{mn}b_{Mn}\gamma_{mn}{}^2\gamma_{Mn}{}^2}\left(-\dfrac{p_b^4}{8m^3b_{mn}{}^3c^2}-\dfrac{p_b^2}{8M^3b_{Mn}{}^3c^2}+U_b\left(\boldsymbol{p}_{bm},\boldsymbol{p}_{bM},\boldsymbol{r}'\right)\right)\end{array}\right]\psi(\boldsymbol{r}')=W'\psi(\boldsymbol{r}') \qquad (10)$$

where

$$W'=W/\left(b_{mn}b_{Mn}\gamma_{mn}{}^2\gamma_{Mn}{}^2\right). \qquad (11)$$

The obtained eq. (10) completed by the expressions (8), (9), (11) represents the basic equation for the quantum two-body problem within the framework of PBFT. Here one should recall that eq. (10) itself, like the source Breit equation, is semi-relativistic, and it is valid to the order $(Z\alpha)^4$. At the same time, the factors $b_{mn}$, $b_{Mn}$, $\gamma_{mn}$ and $\gamma_{Mn}$, being explicitly determined to the orders $(Z\alpha)^2$ and $(Z\alpha)^4$ (see below), allow us to analyze the specific PBFT corrections to the order $(Z\alpha)^6$, which corresponds to the scale of hyperfine interactions. The determination of these corrections is the next goal of our analysis, but, first of all, let us show that eq. (10) yields the same gross and fine structure of the atomic energy levels, as the one furnished by the common approach.

In the zeroth approximation, when the terms of order $(v/c)^2$ and higher are ignored, we get from eq. (10) the Schrödinger equation expressed in $\boldsymbol{r}'$-coordinates:

$$\left(-\frac{\hbar^2\nabla_{r'}^2}{2m_R}-\frac{Ze^2}{r'}\right)\psi(\boldsymbol{r}')=W\psi(\boldsymbol{r}'),$$

where $m_R=mM/(m+M)$ is the reduced mass. Hence we obtain the well-known solution

$$W_{0n}=-\frac{m_Rc^2(Z\alpha)^2}{2n^2},$$

along with the common Schrödinger wave function expressed via $\boldsymbol{r}'$-coordinates. This result allows us to obtain the coefficients $b_{mn}$, $b_{Mn}$, $\gamma_{mn}$, $\gamma_{Mn}$ at least to the order $(Z\alpha)^2$, based on their respective classical limits (4a-d), taking into account the known relationships (*e.g.* [12]):

$$\frac{\overline{U}}{mc^2}=-\frac{Ze^2}{\overline{r}\,mc^2}=-\frac{Ze^2}{\overline{r}\,m_Rc^2}\frac{M}{M+m}=-\frac{(Z\alpha)^2}{n^2}\frac{M}{M+m}, \qquad (12a)$$

$$\frac{\overline{U}}{Mc^2}=-\frac{Ze^2}{\overline{r}\,Mc^2}=-\frac{Ze^2}{\overline{r}\,m_Rc^2}\frac{m}{M+m}=-\frac{(Z\alpha)^2}{n^2}\frac{m}{M+m}, \qquad (12b)$$

$$\frac{\overline{v_m{}^2}}{c^2}=\frac{\overline{v_R{}^2}}{c^2}\frac{M^2}{(M+m)^2}=\frac{(Z\alpha)^2}{n^2}\frac{M^2}{(M+m)^2},\quad \frac{\overline{v_M{}^2}}{c^2}=\frac{\overline{v_R{}^2}}{c^2}\frac{m^2}{(M+m)^2}=\frac{(Z\alpha)^2}{n^2}\frac{m^2}{(M+m)^2}, \qquad (12c\text{-}d)$$

where $v_R$ is the reduced velocity. Hence, via the comparison of eqs. (12a-d) with eqs. (4a-d), we obtain the factors $b_{mn}$, $b_{Mn}$, $\gamma_{mn}$, $\gamma_{Mn}$ to the accuracy $(Z\alpha)^2$ as follows:

$$b_{mn}=\left(1-\frac{(Z\alpha)^2}{n^2}\frac{M}{M+m}\right),\ b_{Mn}=\left(1-\frac{(Z\alpha)^2}{n^2}\frac{m}{M+m}\right) \qquad \text{(13a-b)}$$

$$\gamma_{mn}=\left[1-\frac{(Z\alpha)^2}{n^2}\frac{M^2}{(m+M)^2}\right]^{-1/2},\ \gamma_{Mn}=\left[1-\frac{(Z\alpha)^2}{n^2}\frac{m^2}{(m+M)^2}\right]^{-1/2}. \qquad \text{(13c-d)}$$

Further on, using eqs. (13a-d), we derive the product

$$b_{mn}b_{Mn}\gamma_{mn}{}^2\gamma_{Mn}{}^2=1-\frac{(Z\alpha)^2}{n^2}\frac{2mM}{(M+m)^2} \qquad (14)$$

to the accuracy of calculations $(Z\alpha)^2$.

Applying equations (13), (14), as well as the equality [11]

$$\overline{p^2}=2m_R\overline{\left(W+\frac{Ze^2}{r}\right)}=m_R{}^2c^2(Z\alpha)^2/n^2\ , \qquad (15)$$

we find that to the accuracy of calculations $(Z\alpha)^4$

$$\frac{p_b^2(r')b_{Mn}}{2m}+\frac{p_b^2(r')b_{mn}}{2M}-\frac{W}{b_{mn}b_{Mn}\gamma_{mn}{}^2\gamma_{Mn}{}^2}=\frac{p^2(r')}{2m}-W\ . \qquad (16)$$

Substituting this equality into eq. (10), and ignoring the PBFT factors $b_{mn}$, $b_{Mn}$, $\gamma_{mn}$, $\gamma_{Mn}$ in the terms of the order $(Z\alpha)^4$, we obtain:

$$\left[-\frac{\hbar^2\nabla_{r'}^2}{2m_R}-\frac{Ze^2}{r'}+\left(-\frac{p^4}{8m^3c^2}-\frac{p^4}{8M^3c^2}+U(\boldsymbol{p}_m,\boldsymbol{p}_M,\boldsymbol{r}')\right)\right]\psi(\boldsymbol{r}')=W\psi(\boldsymbol{r}'), \qquad (17)$$

where the term $U(\boldsymbol{p}_m,\boldsymbol{p}_M,\boldsymbol{r}')$ differs from the term (8) by the omission of PBFT factors $b_{mn}$, $b_{Mn}$, $\gamma_{mn}$, $\gamma_{Mn}$.

Excluding further the term of spin-spin interaction in the expression for $U(\boldsymbol{p}_m,\boldsymbol{p}_M,\boldsymbol{r}')$ (last term in the *rhs* of eq. (8)), we arrive at the common solution for the Dirac-Recoil (DR) contribution to the energy levels, written to the order $(Z\alpha)^4$ [3]:

$$\left(W_b^{DR}\right)_{nlj}=m_Rc^2\left\{[f(n,j)-1]-\frac{m_R}{2(m+M)}[f(n,j)-1]^2\right\}, \qquad (18)$$

$$\text{where}\quad f(n,j)\approx 1-\frac{(Z\alpha)^2}{2n^2}-\frac{(Z\alpha)^4}{2n^3}\left(\frac{1}{j+1/2}-\frac{3}{4n}\right),$$

and $j$ is the quantum number of total angular momentum ($\boldsymbol{j}=\boldsymbol{l}+\boldsymbol{s}$, $\boldsymbol{l}$ is the angular momentum, and $\boldsymbol{s}$ the electron's spin).

Thus the corrections of BPFT to the common solutions of equations of atomic physics may emerge at least in the order $(Z\alpha)^6$, which corresponds to the scale of hyperfine interactions, and which will be determined in the next section. As we will see below, these corrections considerably eliminate the available discrepancy between theory and experiment in physics of light hydrogenlike atoms.

# 3. Hyperfine contributions to the atomic energy levels: Pure Bound Field theoretical approach

We have shown above that the solution of the modified Breit equation gives the same gross as well as fine structure of energy levels for hydrogenlike atoms, as that yield by the conventional solution, up to the order of $(Z\alpha)^4$. In this section we analyze hyperfine contributions to the atomic energy levels in PBFT, which can be presented in the form

$$(W_b)_{nlj} = (W_b^{DR})_{nlj} + (W_b)_{HFS} + (L_b)_{nlj}, \qquad (19)$$

where $(W_b)_{HFS}$ is the hyperfine splitting due to spin-spin interaction, and $(L_b)_{nlj}$ is the Lamb shift. Herein the subscript "*b*" reminds that the corresponding energy term is evaluated within the PBFT framework.

In what follows, we consequently analyze the fine structure corrections (sub-section 3.1), corrections to spin-spin interaction (sub-section 3.2), and corrections to the Lamb shift in light hydrogenlike atoms (sub-section 3.3) within the approach of PBFT.

## 3.1. Corrections to the fine structure

In this sub-section we show that the corrections of PBFT to fine structure of light hydrogenlike atoms with $m<<M$ have the order of magnitude of $(Z\alpha)^6 \dfrac{m}{M}$ and higher, and due to their scaling as $n^{-6}$ (or $n^{-5}$), they should be taken into account practically only for the ground states. In the case of positronium, the fine structure corrections occur significant not only due to the equality $m=M$, but also due to the PBFT correction to the annihilation term.

First of all, we extract from the Hamiltonian of eq. (10) the terms of fine interactions with the order $(Z\alpha)^4$ properly modified in PBFT, omitting at this stage the contribution due to spin-spin interaction. Hence, via eqs. (8) and (10) we obtain the operator of fine interaction in the form

$$(V_b(\boldsymbol{r}))_{fine} = (V_b(\boldsymbol{r}))_{rel} + (V_b(\boldsymbol{r}))_{contact} + (V_b(\boldsymbol{r}))_{s-o}, \qquad (20)$$

where

$$(V_b(\boldsymbol{r}))_{rel} = \frac{1}{b_{mn} b_{Mn} \gamma_{mn}^{\ 2} \gamma_{Mn}^{\ 2}} \left( -\frac{p_b^4}{8m^3 b_{mn}^{\ 3} c^2} - \frac{p_b^4}{8M^3 b_{Mn}^{\ 3} c^2} \right) \qquad (21)$$

is the relativistic term,

$$(V_b(\boldsymbol{r}))_{contact} = -\frac{1}{b_{mn} b_{Mn} \gamma_{mn}^{\ 2} \gamma_{Mn}^{\ 2}} \frac{\pi Z e^2 \hbar^2}{2c^2} \left( \frac{1}{m^2 b_{mn}^{\ 2}} + \frac{1}{M^2 b_{Mn}^{\ 2}} \right) \delta(\boldsymbol{r}) \qquad (22)$$

is the term of contact interaction, and

$$\left(V_b(\boldsymbol{r})\right)_{s-o}=-\frac{Ze^2\hbar}{b_{mn}b_{Mn}\gamma_{mn}\gamma_{Mn}c^2r^3}\left[\frac{(\boldsymbol{r}\times\boldsymbol{p}_{bm})\cdot\sigma_m}{2m^2{b_{mn}}^2}+\frac{(\boldsymbol{r}\times\boldsymbol{p}_{bM})\cdot\sigma_M}{2M^2{b_{Mn}}^2}-\frac{((\boldsymbol{r}\times\boldsymbol{p}_{bm})\cdot\sigma_M-(\boldsymbol{r}\times\boldsymbol{p}_{bM})\cdot\sigma_m)}{2b_{mn}b_{Mn}mM}\right] \quad (23)$$

is the term of spin-orbit interaction.

We point out that the operators (21)-(23) are presented in $r$-coordinates, whereas in the Hamiltonian (10) they should be expressed through $r'$– coordinates. In the latter case, each of the terms (21)-(23) can be calculated with taking into account of eq. (9), as well as the equalities

$$p_b^4(\boldsymbol{r}')={b_{mn}}^4{b_{Mn}}^4{\gamma_{mn}}^4{\gamma_{Mn}}^4p_b^4(\boldsymbol{r}),\ \delta(\boldsymbol{r})={b_{mn}}^3{b_{Mn}}^3{\gamma_{mn}}^3{\gamma_{Mn}}^3\delta(\boldsymbol{r}'). \quad (24\text{-}25)$$

Introducing the replacements (24-25) into eq. (21-23), and expressing the terms (22), (23) via the fine structure constant, we obtain the operator of fine interactions in $\boldsymbol{r}'$-coordinates as follows:

$$\left(V_b(\boldsymbol{r}')\right)_{fine}={\gamma_{mn}}^2{\gamma_{Mn}}^2\left(\begin{array}{l}-\frac{p_b^4(\boldsymbol{r}'){b_{Mn}}^3}{8m^3c^2}-\frac{p_b^4(\boldsymbol{r}'){b_{mn}}^3}{8M^3c^2}+\frac{\pi(Z\alpha){b_{Mn}}^2}{2m^2c^2}\delta(\boldsymbol{r}')+\frac{\pi(Z\alpha){b_{mn}}^2}{2M^2c^2}\delta(\boldsymbol{r}')+\\ \frac{(Z\alpha){b_{Mn}}^2}{2m^2r'^3c^2}(\boldsymbol{r}'\times\boldsymbol{p}_{bm}(r'))\cdot\sigma_m+\frac{(Z\alpha){b_{mn}}^2}{2M^2r'^3c^2}(\boldsymbol{r}'\times\boldsymbol{p}_{bM}(r'))\cdot\sigma_M-\\ \frac{(Z\alpha)b_{mn}b_{Mn}}{2mMc^2r^3}((\boldsymbol{r}\times\boldsymbol{p}_{bm})\cdot\sigma_M-(\boldsymbol{r}\times\boldsymbol{p}_{bM})\cdot\sigma_m)\end{array}\right). \quad (26)$$

Substituting this value in eq. (10), we get

$$\left[-\frac{\hbar^2\nabla_{r'}^2b_{Mn}}{2m}-\frac{\hbar^2\nabla_{r'}^2b_{mn}}{2M}-\frac{Ze^2}{r'}+\left(V_b(\boldsymbol{r}')\right)_{fine}\right]\psi(\boldsymbol{r}')=\frac{W}{b_{mn}b_{Mn}{\gamma_{mn}}^2{\gamma_{Mn}}^2}\psi(\boldsymbol{r}'). \quad (27)$$

We emphasize that to the order $(Z\alpha)^4$, eq. (27) is equivalent to eq. (17), which yields the same fine structure of energy levels, like the common approach. Now our goal is to determine the fine structure corrections of PBFT in eq. (27), which may emerge in the order $(Z\alpha)^6$. To solve this problem, we need to determine factors $b_{mn}$, $b_{Mn}$, $\gamma_{mn}$, $\gamma_{Mn}$ to the order $(Z\alpha)^4$.

For this purpose, first of all, we find the classical coefficients $b_m$, $b_M$, and $\gamma_m$, $\gamma_M$ to the accuracy $c^{-4}$ based on their definitions (4a-d). Concerning the factors $b_m$, $b_M$, their modification issues from the two circumstances:

- the factors $\gamma_m$ and $\gamma_M$ in eqs. (4a-b) are no longer adopted to be equal to unity, but they have to be determined to the accuracy $(v/c)^2$;
- in the expression for interaction EM energy (2), the term of magnetic interaction energy is no longer ignored. Taking into account that for the bound EM field $\boldsymbol{B}=(\boldsymbol{v}\times\boldsymbol{E})/c$, we obtain for the circular motion in the classical two-body problem

$$\boldsymbol{B}_m\cdot\boldsymbol{B}_M=B_mB_M=\frac{v_mv_M}{c^2}E_mE_M=\frac{mM}{(m+M)^2}\frac{{v_R}^2}{c^2}\boldsymbol{E}_m\cdot\boldsymbol{E}_M.$$

Substituting this equality into eq. (2) and following the quantum mechanical definition of $\overline{U}$, we obtain

$$b_{mn}=\left(1+\frac{\gamma_{Mn}\overline{U}}{mc^2}\right)=1-\frac{(Z\alpha)^2}{n^2}\left(1+\frac{(Z\alpha)^2}{2n^2}\frac{m^2}{(m+M)^2}\right)\left(1+\frac{mM}{(m+M)^2}\frac{(Z\alpha)^2}{n^2}\right)\frac{M}{(m+M)}\approx$$

$$1-\frac{(Z\alpha)^2}{n^2}\frac{M}{m+M}-\frac{(Z\alpha)^4}{2n^4}\frac{m^2M}{(m+M)^3}-\frac{(Z\alpha)^4}{n^4}\frac{mM^2}{(m+M)^3} \qquad (28a)$$

to the accuracy $(Z\alpha)^4$, where we have used eq. (13d), putting

$$\left[1-\frac{(Z\alpha)^2}{n^2}\frac{m^2}{(m+M)^2}\right]^{-1/2}\approx 1+\frac{(Z\alpha)^2}{2n^2}\frac{m^2}{(m+M)^2}$$

with the sufficient accuracy of calculations.

Similarly we determine the factor

$$b_{Mn}=1-\frac{(Z\alpha)^2}{n^2}\frac{m}{m+M}-\frac{(Z\alpha)^4}{2n^4}\frac{M^2m}{(m+M)^3}-\frac{(Z\alpha)^4}{n^4}\frac{Mm^2}{(m+M)^3}. \qquad (28b)$$

We point out that expanding the classical coefficients $b_{mn}$ and $b_{Mn}$ (eqs. (4a-b)) to the accuracy $(v/c)^4$, we adopt that the involvement of non-Coulomb interactions does not affect their values. Indeed, at the semi-classical level, as shown in ref. [13], such non-Coulomb interactions are exactly counteracted by corresponding change of the Coulomb interaction energy due to the proper variation of the radius of electron's orbit. Hence the resultant action of Coulomb and non-Coulomb interactions is equivalent to the Coulomb interaction alone with the fixed electron's orbit. This observation concurrently implies that the overall change of the energy of semi-classical system "electron plus nucleus", for example, due to spin-orbit interaction exhibits as a proper change of kinetic energy of the orbiting electron and nucleus by the energy of non-Coulomb interaction [13]. Therefore, the taking into account of fine interactions does modify the factors $\gamma_{mn}$ $\gamma_{Mn}$ in the orders $(Z\alpha)^4$ and higher. In order to find the related corrections, it is convenient to use the relationships between the Lorentz factors and momenta of particles

$$\gamma_{mn}{}^2=\frac{1}{1-p_b^2/m^2c^2},\ \gamma_{Mn}{}^2=\frac{1}{1-p_b^2/M^2c^2},$$

where $p_b$ is the value of non-relativistic momentum of electron (nucleus), and to the order $(Z\alpha)^2$ the averaged value of $p_b^2$ is defined by eq. (15).

With the involvement of fine interactions, this expression is modified as

$$\overline{p^2}=2m_R\left(\overline{\left(W+V_{fine}\right)+\frac{Ze^2}{r}}\right).$$

Taking into account that $\overline{W+\frac{Ze^2}{r}}=mc^2(Z\alpha)^2$ [11], we obtain

$$\gamma_{mn}{}^2=\frac{1}{1-\frac{(Z\alpha)^2}{n^2}\frac{M^2}{(m+M)^2}-\frac{2m_R(V_b(\mathbf{r}'))_{fine}}{m^2c^2}},\ \gamma_{Mn}{}^2=\frac{1}{1-\frac{(Z\alpha)^2}{n^2}\frac{m^2}{(m+M)^2}-\frac{2m_R(V_b(\mathbf{r}'))_{fine}}{M^2c^2}}.\quad\text{(28c-d)}$$

Having obtained the coefficients $b_{mn}$ and $b_{Mn}$ (eqs. (28a-b)) and $\gamma_{mn}$, $\gamma_{Mn}$ (eqs. (28c-d) we now in the position to calculate the fine structure corrections of PBFT to the order $(Z\alpha)^6$ on the basis of eq. (27). In particular, we find that

$$\frac{p^2b_{Mn}}{2m}+\frac{p^2b_{mn}}{2M}=\frac{p^2}{2m_R}-m_Rc^2\frac{(Z\alpha)^4}{n^4}\frac{Mm}{(m+M)^2}-(V_b(\mathbf{r}'))_{fine}\frac{(Z\alpha)^2}{n^2}\frac{2Mm}{(m+M)^2}. \quad (29)$$

Further, we determine the product $\frac{W}{b_{mn}b_{Mn}\gamma_{mn}{}^2\gamma_{Mn}{}^2}$, which for the $n$S state is equal to

$$\frac{W}{b_{mn}b_{Mn}\gamma_{mn}{}^2\gamma_{Mn}{}^2}=W-\frac{m_Rc^2(Z\alpha)^4}{n^4}\frac{Mm}{(m+M)^2}+\overline{V}_f\frac{(Z\alpha)^2}{n^2}-\frac{m_Rc^2(Z\alpha)^6 5mM}{4n^6(m+M)^2}-\frac{m_Rc^2(Z\alpha)^6m^2M^2}{2n^6(m+M)^4}. \quad (30)$$

Here we have used eqs. (28a-d), and in the terms of the order $(Z\alpha)^4$ we put $W$=$W_{0n}$=$-\frac{m_Rc^2(Z\alpha)^2}{2n^2}$,

while in the terms of the lower order $(Z\alpha)^2$ we presented $W=W_{0n}+\overline{V}_{fine}$.

In order to find the PBFT corrections for the term $(V_b(\mathbf{r}'))_{fine}$, we can write it in the form

$$(V_b(\mathbf{r}'))_{fine}=(V(\mathbf{r}'))_{fine}+\delta V_{fine},$$

where $(V(\mathbf{r}'))_{fine}$ represents the common operator of fine structure, and $\delta V_{fine}$ contains the specific corrections of PBFT determined by the substitution of factors $b_{mn}$, $b_{Mn}$, $\gamma_{mn}$, $\gamma_{Mn}$ in eq. (26). Since the operator $(V(\mathbf{r}'))_{fine}$ itself has the order $(Z\alpha)^4$, it is enough to use the PBFT factors (13a-d) written to the accuracy $(Z\alpha)^2$. In particular, for $nS$-state of hydrogenlike atom, the straightforward calculations yield:

$$\delta V_{fine}=\frac{M^2+m^2}{(M+m)^2}\frac{(Z\alpha)^2}{n^2}\overline{V}_f+\frac{2mc^2(Z\alpha)^6}{n^5(M+m)^5}\left(mM^4-m^2M^3+m^3M^2\right)-\frac{9}{8}\frac{mc^2(Z\alpha)^6}{n^6}\frac{mM^2(M^2+m^2)}{(M+m)^5}. \quad (31)$$

Substituting eqs. (28-31) into eq. (27), we obtain after lengthy, but straightforward calculations for $nS$-states:

$$\left[\frac{p^2}{2m_R}-\frac{Ze^2}{r'}+(V)_f\right]\psi(\mathbf{r}')=$$
$$\left\{\begin{aligned}&W-\frac{2mc^2(Z\alpha)^6}{n^5(M+m)^5}\left(2mM^4+m^2M^3+2m^3M^2\right)+\frac{mc^2(Z\alpha)^6}{n^6}\frac{mM^2}{4(m+M)^3}+\frac{m_Rc^2(Z\alpha)^6m^2M^2}{2n^6(m+M)^4}\\&+mc^2\frac{(Z\alpha)^6}{2n^6}\frac{(M^2+m^2)mM^2}{(m+M)^5}+\frac{9}{8}\frac{mc^2(Z\alpha)^6}{n^6}\frac{mM^2(M^2+m^2)}{(M+m)^5}\end{aligned}\right\}\psi(\mathbf{r}') \quad (32)$$

Designating the term in the bracket of *rhs* of eq. (32) as $W_{common}$, we rewrite this equation in the short form

$$\left[\frac{p^2}{2m_R}-\frac{Ze^2}{r'}+(V)_f\right]\psi(\boldsymbol{r}')=W_{common}\psi(\boldsymbol{r}'),$$

which represents the common Breit equation without spin-spin term written in the Schrödinger-like form (see, *e.g.* [11]). Therefore, the PBFT correction to the atomic *nS* state energy levels is equal to

$$\delta W_{PBFT}=W-W_{common}=\frac{2mc^2(Z\alpha)^6}{n^5(M+m)^5}\left(2mM^4+m^2M^3+2m^3M^2\right)-$$
$$\frac{mc^2(Z\alpha)^6}{n^6}\frac{mM^2}{4(m+M)^3}-\frac{m_Rc^2(Z\alpha)^6m^2M^2}{2n^6(m+M)^4}-mc^2\frac{(Z\alpha)^6}{2n^6}\frac{(M^2+m^2)mM^2}{(m+M)^5}-\frac{9}{8}\frac{mc^2(Z\alpha)^6}{n^6}\frac{mM^2(M^2+m^2)}{(M+m)^5}$$
(33)

One can see that for atoms with $m<<M$, the correction (33) has the order of magnitude $mc^2(Z\alpha)^6 m/M$ and scales as $n^{-6}$ or $n^{-5}$. Therefore, there is no practical need to calculate this correction for $n\geq 2$ and $l\neq 0$.

In fact, the correction (33) is significant for 1*S* state only, and for the hydrogen atom it is equal to

$$\delta W_{fine}^{\mathrm{H}}(1S)=\frac{13}{8}mc^2(Z\alpha)^6\frac{m}{M}=14.8\ \text{kHz} \qquad (34a)$$

Note that this correction is positive, and it reduce the value of 1*S* state (which is negative). For 2*S* state of hydrogen we obtain from eq. (31) $\delta W_{fine}^{\mathrm{H}}(2S)$ =0.91 kHz. Hence the correction to 1*S*-2*S* interval is equal to

$$\delta W_{fine}^{\mathrm{H}}(1S-2S)=13.9\ \text{kHz}. \qquad (34b)$$

Below the correction (34b) will be involved into the re-estimation of 1*S* Lamb shift in hydrogen.

For 1*S* state of muonium, this correction is about 0.13 MHz, and is few times less than the theoretical uncertainty in calculation of 1*S*-2*S* transition (≈0.30 MHz [3]), and much less than the experimental uncertainty in the measurement of this interval (9.8 MHz [14]).

For 1*S* state of positronium (*m*=*M*), the correction (33) becomes

$$\delta W_{fine}^{\mathrm{Ps}}(1S)=\frac{21mc^2(Z\alpha)^6}{128}=3.07\ \text{MHz},$$

and for the 1*S*-2*S* interval, as follows from eq. (33),

$$\delta W_{fine}^{\mathrm{Ps}}(1S-2S)=2.95\ \text{MHz}, \qquad (34c)$$

which will be involved below into the re-estimation of 1*S*-2*S* interval in positronium.

Besides, for positronium the Breit potential includes the additional annihilation part (*e.g.*, [10, 11]), which in PBFT acquires the form

$$\left(V_b(\boldsymbol{r})\right)_{ann}=\frac{1}{b_n^{\ 2}\gamma_n^{\ 4}}\left(\frac{\pi(Z\alpha)}{2m^2b_n^{\ 2}c^2}\left(3+\ \boldsymbol{\sigma}_+\cdot\ \boldsymbol{\sigma}_-\right)\delta(\boldsymbol{r})\right),$$

where $\boldsymbol{\sigma}_+$ ( $\boldsymbol{\sigma}_-$ ) belongs to positron (electron), and we designated $b_{mn}=b_{Mn}=b_n$, $\gamma_{mn}=\gamma_{Mn}=\gamma_n$. In $r'$-coordinates, due to eq. (25), this operator acquires the form

$$\left(V_b(\boldsymbol{r}')\right)_{ann}=b_n^{\ 2}\gamma_n^{\ 2}\left(\frac{\pi(Z\alpha)}{2m^2c^2}\left(3+\ \boldsymbol{\sigma}_+\cdot\ \boldsymbol{\sigma}_-\right)\delta(\boldsymbol{r}')\right).$$

We average this term with the wave function for $l$=0 [11]

$$\psi(0)=\frac{1}{\sqrt{\pi}}\left(\frac{Z\alpha m_R}{n^2}\right)^{3/2},$$

and taking into account that for positronium $b_n=\left(1-\frac{(Z\alpha)^2}{2n^2}\right)$, $\gamma_{mn}=\left[1-\frac{(Z\alpha)^2}{4n^2}\right]^{-1/2}$ (see eqs. (13) for $m$=$M$), we derive in the orthopositronium case:

$$\left(W_b\right)_{ann}=\left(\frac{mc^2(Z\alpha)^4}{4n^3}\left(1-\frac{3(Z\alpha)^2}{4}\right)\right).$$

Hence the correction of PBFT to annihilation term reads:

$$\delta W_{ann}=\frac{3mc^2(Z\alpha)^6}{16n^5}. \qquad (35)$$

The correction (35) decreases the value of 1$S$-2$S$ interval in positronium by 3.40 MHz, and thus, it should be added to the fine structure correction (34c). A comparison of the theory and experiment for 1$S$-2$S$ interval in positronium will be done below in sub-section 4.1.

### 3.2. Corrections to hyperfine splitting of energy levels due to spin-spin interaction

Now we analyze the contribution of spin-spin interaction into the Breit potential, which in PBFT has the form

$$\left(V_b(\boldsymbol{r})\right)_{s-s}=\frac{1}{b_{mn}b_{Mn}\gamma_{mn}^{\ 2}\gamma_{Mn}^{\ 2}}\frac{Ze^2h^2\gamma_{mn}\gamma_{Mn}}{4mb_{mn}Mb_{Mn}c^2}\left(\frac{\sigma_m\cdot\sigma_M}{r^3}-3\frac{(\sigma_m\cdot\boldsymbol{r})(\sigma_M\cdot\boldsymbol{r})}{r^5}-\frac{8\pi}{3}\delta(\boldsymbol{r})\right) \qquad (36)$$

(the last term of eq. (8)). Being expressed via $r'$-coordinates, this operator reads:

$$\left(V_b(\boldsymbol{r}')\right)_{s-s}=b_{mn}b_{Mn}\gamma_{mn}^{\ 2}\gamma_{Mn}^{\ 2}\frac{e^2h^2}{4mMc^2}\left(\frac{\sigma_m\cdot\sigma_M}{r'^3}-3\frac{(\sigma_m\cdot\boldsymbol{r}')(\sigma_M\cdot\boldsymbol{r}')}{r'^5}-\frac{8\pi}{3}\delta(\boldsymbol{r}')\right), \qquad (37)$$

where we have used eqs. (9) and (25). Designating

$$\left(V(\boldsymbol{r}')\right)_{s-s}=\frac{e^2h^2}{4mMc^2}\left(\frac{\sigma_m\cdot\sigma_M}{r'^3}-3\frac{(\sigma_m\cdot\boldsymbol{r}')(\sigma_M\cdot\boldsymbol{r}')}{r'^5}-\frac{8\pi}{3}\delta(\boldsymbol{r}')\right)$$

(the common Hamiltonian of spin-spin interaction expressed via $r'$-coordinates), and substituting expressions (13a-d) for PBFT factors, determined with the sufficient accuracy $(Z\alpha)^2$, we obtain

$$\left(V_b(\boldsymbol{r}')\right)_{s-s} = \left(1 - \frac{(Z\alpha)^2}{n^2}\frac{2mM}{(M+m)^2}\right)\left(V(\boldsymbol{r}')\right)_{s-s}. \qquad (38a)$$

This relationship is also valid for the energy of spin-spin interaction, obtained via the averaging of operators $\left(V_b(\boldsymbol{r}')\right)_{s-s}$ and $\left(V(\boldsymbol{r}')\right)_{s-s}$ with the wave function $\psi(\boldsymbol{r}')$:

$$\left(W_b\right)_{s-s} = \left(1 - \frac{(Z\alpha)^2}{n^2}\frac{2mM}{(M+m)^2}\right)W_{s-s}. \qquad (38b)$$

Now it is important to remind that the energy $W_{s\text{-}s}$ contains the ratios of magnetic moment to mass both for the electron and the nucleus, which are determined experimentally by means of the Zeeman effect. Since in PBFT the operator of interaction of magnetic dipole with an external magnetic field is, in general, modified, the appropriate corrections to the measured values "magnetic moment/mass" ratio should be clarified, too.

As known, the operator of interaction of two bound particles (electron and nucleus) with the external magnetic field reads [1]:

$$V_{mag} = g_m \frac{e\hbar}{2m}\left(\boldsymbol{s}_m \cdot \boldsymbol{B}\right) - g_M \frac{Ze\hbar}{2M}\left(\boldsymbol{s}_M \cdot \boldsymbol{B}\right), \qquad (39)$$

where $g_m$, $g_M$ are the g-factors for bound electron and nucleus, correspondingly. Being added to the Breit operator of eq. (10) with the PBFT corrections $m \to b_{mn}m$, $M \to b_{mn}M$, $\boldsymbol{B} \to \gamma_{mn}\gamma_{Mn}\boldsymbol{B}$, this operator acquires the form:

$$\left(V_b\right)_{mag} = \frac{1}{b_{mn}b_{Mn}\gamma_{mn}^{\ 2}\gamma_{Mn}^{\ 2}}\left[g_m \frac{e\hbar}{2mb_{mn}}\gamma_{mn}\gamma_{Mn}\left(\boldsymbol{s}_m \cdot \boldsymbol{B}\right) - g_M \frac{Ze\hbar}{2Mb_{Mn}}\gamma_{mn}\gamma_{Mn}\left(\boldsymbol{s}_M \cdot \boldsymbol{B}\right)\right].$$

Averaging this operator with the Schrödinger wave-function $\psi(\boldsymbol{r})$, due to the normalization requirement $\psi(\boldsymbol{r}) = \left(b_{mn}b_{Mn}\gamma_{mn}\gamma_{Mn}\right)^{3/2}\psi(\boldsymbol{r}')$ implied by eq. (9), we obtain

$$\left(\overline{V}_b\right)_{mag} \equiv \left(W_b\right)_{mag} = b_{mn}b_{Mn}\gamma_{mn}^{\ 2}\gamma_{Mn}^{\ 2}\left[g_m \frac{e\hbar b_{Mn}}{2m}\left(\boldsymbol{s}_m \cdot \boldsymbol{B}\right) - g_M \frac{Ze\hbar b_{mn}}{2M}\left(\boldsymbol{s}_M \cdot \boldsymbol{B}\right)\right], \qquad (40)$$

where $\left(W_b\right)_{mag}$ gives the Zeeman splitting of energy levels in PBFT framework. Herein in averaging of $\left(\overline{V}_b\right)_{mag}$ we put $\boldsymbol{B}(\boldsymbol{r}) = const$ within the atomic scale, which is always fulfilled in practice.

Inserting eqs. (13a-d) into eq. (40) and putting $Z$=1 (which is enough for our immediate purposes), we obtain for $nS$-state

$$\left(W_b\right)_{mag} \approx \left(1 - \frac{(Z\alpha)^2}{n^2}\frac{2Mm}{(m+M)^2}\right)\left[W_{mag} - \frac{(Z\alpha)^2}{n^2}\frac{e\hbar}{2m}\left(\left(g_m\boldsymbol{s}_m - g_M\boldsymbol{s}_M\right)\cdot\boldsymbol{B}\right)\frac{1}{M+m}\right], \qquad (41)$$

where $W_{mag}$ stands for the Zeeman splitting of energy levels, obtained via the averaging of common operator (39). For the sublevel $F$=1 used for the measurement of Zeeman effect, $\boldsymbol{s}_m - \boldsymbol{s}_M$ =0, and eq. (41) acquires the form

$$\left(W_b\right)_{mag} \approx \left(1-\frac{(Z\alpha)^2}{n^2}\frac{2mM_Z}{\left(m+M_Z\right)^2}\right)\left(W_{mag}-\frac{(Z\alpha)^2}{n^2}\frac{\left(g_m-g_M\right)e\hbar}{2m}\left(\boldsymbol{s}_m\cdot\boldsymbol{B}\right)\frac{1}{M+m}\right), \tag{42}$$

where we supply the mass $M$ by the subscript "Z" ("Zeeman effect"), in order to distinguish it from the mass $M$ in eqs. (38a-b), designating the mass of the nucleus in the measurement of spin-spin splitting. Thus, the ratio of "magnetic moment/mass" derived from the Zeeman splitting should be corrected in PBFT for each bound particle with taking into account of the relationship (42) between $\left(W_b\right)_{mag}$ and $W_{mag}$.

For the 1$S$ state of hydrogen and heavier atoms, the analysis of this corrections in the estimation of spin-spin interval (38b) is not practically important, because the term $(Z\alpha)^2\frac{2mM}{(M+m)^2}W_{s-s}$ of eq. (38b) itself is less than 100 Hz and is many times smaller than the nuclear-structure corrections to the 1$S$ hyperfine splitting, which vary from tens to hundreds kHz [15-17]. Thus for hydrogen the PBFT correction to spin-spin splitting occurs negligible, and we put

$$\left(W_b\right)_{s-s}^{\mathrm{H}} = W_{s-s}^{\mathrm{H}}. \tag{43}$$

within the range of the present uncertainty in calculation of $W_{s-s}^{\mathrm{H}}$.

Considering the spin-spin splitting of 1$S$ state of muonium, we can also ignore the correction to the ratio "magnetic moment/mass" for the electron, because one can show that it induces the PBFT correction of the order 100 Hz, which is much smaller than the present theoretical uncertainly in calculation of spin-spin interval in muonium (about 500 Hz [1]). Further, for muonium we can put with a high accuracy $g_m$=$g_M$ in the term containing $(Z\alpha)^2$. We also use the known fact that the ratio "magnetic moment/mass" for bound muon is determined with the best accuracy via the Zeeman effect in muonium [1], so $M_Z$=$M$. With these equalities eq. (42) yields

$$\left(W_b\right)_{mag}^{\mathrm{Mu}} = \left(1-\frac{(Z\alpha)^2}{n^2}\frac{2mM}{(m+M)^2}\right)W_{mag}^{\mathrm{Mu}}, \tag{44}$$

which shows that the PBFT correction to spin-spin interval (38b) has exactly the same structure as PBFT correction to the "magnetic moment/mass" ratio for bound muon (44). As a result, both corrections exactly compensate each other, and we get

$$\left(W_b\right)_{s-s}^{\mathrm{Mu}} = \frac{\left(1-\frac{(Z\alpha)^2}{n^2}\frac{2mM}{(M+m)^2}\right)}{\left(1-\frac{(Z\alpha)^2}{n^2}\frac{2mM_Z}{(m+M_Z)^2}\right)}W_{s-s}^{\mathrm{Mu}} = W_{s-s}^{\mathrm{Mu}}. \tag{45}$$

For positronium we have quite different situation ($M$=$m$, but $M_Z \approx 10^3 m$ if the magnetic moment/mass ratio for electron is taken from the Zeeman splitting in hydrogen or heavier atoms), and the correction to spin-spin interaction (38b) dominates over the correction to the "magnetic moment/mass" ratio for bound electron. Hence we involve the correction of eq. (38b) solely, which yields

$$\left(W_b\right)_{s-s}^{\mathrm{Ps}} = \left(1-\frac{(Z\alpha)^2}{2n^2}\right)W_{s-s}^{\mathrm{Ps}}. \qquad (46)$$

This equation will be used in sub-section 4.2 for PBFT correction of spin-spin interval for 1S state of positronium and for comparison with modern experimental data.

### 3.3. Corrections to the Lamb shift.

The corrections of PBFT obtained above in sub-sections 3.1 and 3.2 originate from the appropriate modification of DC equation suggested in ref. [7]. Analyzing now radiative corrections to the atomic energy levels, one should emphasize that the modifications in the Dirac equation do not touch the core structure of QED, but at the same time imply the replacements (5a-c) in the input of QED expressions. On the basis of this result we derive below the corrections to the Lamb shift $L$ for light hydrogenlike atoms, which emerge in PBFT.

It is known that the dominant terms of the Lamb shift arise due to a finite radius of the electron $\langle r^2\rangle$, which continuously emits and absorbs virtual photons, as well as due to vacuum polarization.

The finite radius of the electron induces a deviation from the Coulomb potential [3]

$$\delta V_{\text{finite radius}} = \frac{1}{6}\langle r^2\rangle \Delta U \approx \frac{\alpha}{3\pi}\ln(Z\alpha)^{-2}\frac{\Delta U}{m^2} = \frac{4}{3}\ln(Z\alpha)^{-2}\frac{Z\alpha^2}{m^2}\delta(\boldsymbol{r}), \qquad (47)$$

where $\Delta$ is the Laplacian. According to eqs. (5a-c), for the bound electron the mass $m$ is replaced by $b_{mn}m$, while the Coulomb potential $U$ is replaced by $U'=\gamma_{mn}\gamma_{Mn}U\approx\gamma_{mn}U$, where we can put $\gamma_{Mn}=1$ with the sufficient accuracy of calculations; further we also put $b_{Mn}=1$ in the correction to the Lamb shift. Thus the latter equation acquires the form

$$\delta V'_{\text{finite radius}} = \frac{4\gamma_n}{3b_n^2}\ln(Z\alpha)^{-2}\frac{Z\alpha^2}{m^2}\delta(\boldsymbol{r}) = \frac{\gamma_{mn}}{b_{mn}^2}\delta V_{\text{finite radius}}, \qquad (48a)$$

where $\delta V_{\text{finite radius}}$ is defined by eq. (47).

The contribution due to vacuum polarization [3] is also proportional to $\Delta U/m^2$, and thus the latter equation remains in force for this correction:

$$\delta V'_{\text{polarization}} = \frac{\gamma_{mn}}{b_{mn}^2}\delta V_{\text{polarization}}. \qquad (48b)$$

The total contribution $\delta V_{\text{total}}$ is defined as the sum $\delta V_{\text{total}} = \delta V_{\text{finite radius}} + \delta V_{\text{polarization}}$, so that for the total perturbation we get

$$\delta V'_{\text{total}} = \frac{\gamma_{mn}}{b_{mn}^2}\delta V_{\text{total}}. \qquad (48c)$$

The correction to the energy level is given by the matrix element of the total perturbation (48c), where we need to take into account that due to normalization requirement implied by the scaling transformation (9), $\psi(\boldsymbol{r}) = (b_{mn}b_{Mn}\gamma_{mn}\gamma_{Mn})^{3/2}\psi(\boldsymbol{r}') \approx b_{mn}^{3/2}\gamma_{mn}^{3/2}\psi(\boldsymbol{r}',\vartheta,\varphi)$. Hence

$$\Delta W_{PBFT} = \langle nS|\delta V'_{total}|nS\rangle = b_{mn}{}^3\gamma_{mn}{}^3\left(\frac{\gamma_{mn}}{b_{mn}{}^2}\Delta E\right) = b_{mn}\gamma_{mn}{}^4\Delta W = \gamma_{mn}{}^2\Delta W ,$$

where $\Delta E$ denotes the value of energy shift, obtained within QED, and we have used the equality $b_{mn}\gamma_{mn}{}^2 = 1$, followed from eq. (14) at $M\to\infty$.

Thus the Lamb shift at the given energy level corrected within PBFT, reads:

$$\left(L_{nlj}\right)_{PBFT} = \gamma_{mn}{}^2 L_{nlj} , \qquad (49)$$

where $L_{nlj}$ stands for the Lamb shift calculated in QED.

For the $2S_{1/2}$-$2P_{1/2}$ Lamb shift both levels have a principal quantum number $n$=2, and we get

$$\left(L_b\right)_{2S-2P} = \gamma_{m2}{}^2 L_{2S-2P} = L_{2S-2P}\left(1-(Z\alpha)^2/4\right)^{-1}. \qquad (50)$$

Thus the correction induced by PBFR to the $2S$-$2P$ Lamb shift is equal to

$$\delta L_{2S-2P} = \left(L_b\right)_{2S-2P} - L_{2S-2P} = L_{2S-2P}\left[\left(1-(Z\alpha)^2/4\right)^{-1} - 1\right]. \qquad (50a)$$

Numerically this value is equal to 13.8 kHz, which exceeds substantially the measured precision for Doppler-free two-photon laser spectroscopy [18].

Below we compare the experimental and theoretical values for the corrected $2S$-$2P$ Lamb shift in hydrogen (sub-section 4.3) and $2S$-$2P$ Lamb shift in $He^+$ (sub-section 4.4).

Eq. (49) is, in general, also applicable to the $1S$ Lamb shift $L_{1S}$ in hydrogenlike atoms. However, its direct measurement is impractical until effects of nuclear structure are known accurately enough. In order to eliminate the influence of these effects, the data at least of two measurements are involved: for hyperfine intervals in the ground state and metastable states (for example, for the 1S and $2S$ states). Since the bulk contribution to the Lamb shift scales like $n^{-3}$, then the difference $8(W_{hpf})_{2S} - (W_{hpf})_{1S}$ allows us canceling substantially various contributions caused by the short distance effects. However, the factors $\gamma_{mn}$ differ from each other for $2S$ and $1S$ states, and calculation of the corrected $1S$ Lamb shift $\left(L_b\right)_{1S}$ is not straightforward.

In order to introduce the PBFT corrections to the 1S Lamb shift, to be convenient for practical applications, one need to look closer at the typical methods for its theoretical estimation. In principle, the $1S$ Lamb shift could be extracted from the experimental data on the transition frequencies between the energy levels with different numbers $n$. One should emphasize that the intervals of gross structure are mainly determined by the Rydberg constant $R$. In order to disentangle measurement of the $1S$ Lamb shift from the measurement of the Rydberg constant, one can use the experimental data on two different intervals $1S$-$2S$ and $2S_{1/2}$-$8D_{5/2}$ of hydrogen [19]. Theoretically these intervals can be presented as [3]

$$E_{1S-2S} = \left(W^{DR}_{2S_{1/2}} - W^{DR}_{1S_{1/2}}\right) + L_{2S_{1/2}} - L_{1S_{1/2}} , \qquad (51a)$$

$$E_{2S-8D} = \left(W^{DR}_{8D_{1/2}} - W^{DR}_{1S_{1/2}}\right) + L_{8D_{5/2}} - L_{2S_{1/2}} , \qquad (51b)$$

where $W^{DR}_{nl_j}$ is the leading Dirac and recoil contribution to the position of the respective energy level (eq. (18)).

The differences of the leading Dirac and recoil contribution in the *rhs* of eq. (51) are proportional to the Rydberg constant $R$ plus corrections of order $\alpha^2 R$ and higher. One can construct a linear combination of these intervals which is proportional to $\alpha^2 R$ plus higher order terms

$$E_{1S-2S}-\frac{16}{5}E_{2S-8D}=\left(W_{2S}^{DR}-W_{1S}^{DR}\right)-\frac{16}{5}\left(W_{8D}^{DR}-W_{2S}^{DR}\right)-L_{1S}+\frac{21}{5}L_{2S}-\frac{16}{5}L_{8D}\,. \qquad (52)$$

Then the difference of the leading Dirac recoil contribution in the *rhs* of eq. (52) can be calculated with a high accuracy, due to the suppression factor $\alpha^2$, and it practically does not depend on the exact value of $R$. Hence the linear combination of the Lamb shifts in the *rhs* of eq. (52) does not depend on $R$, too. The bulk contribution to the Lamb shift scales as $1/n^3$ which allows using the theoretical value $L_{8D_{5/2}}$ =71.51 kHz [3] without loss of accuracy. The 2*S* Lamb shift can be extracted from the data on the classic 2*S*-2*P* Lamb shift, so that

$$L_{1S}=\left[\left(W_{2S}^{DR}-W_{1S}^{DR}\right)-\frac{16}{5}\left(W_{8D}^{DR}-W_{2S}^{DR}\right)-\frac{16}{5}L_{8D}\right]-\left[E_{1S-2S}-\frac{16}{5}E_{2S-8D}\right]+\frac{21}{5}L_{2S}. \qquad (53)$$

Herein in *rhs*, the first term in square brackets is computed theoretically, the second term in square brackets is determined experimentally, while the last term is extracted from the data on the classic 2*S*-2*P* Lamb shift. Within PBFT, the first computed term in the *rhs* of eq. (53) should be corrected by adding the fine structure correction (34b). Besides, one has to correct within PBFT the 2*S* Lamb shift, using the data on the classic 2*S*-2*P* Lamb shift. Hence the expression for the 1*S* Lamb shift for the hydrogen within PBFT acquires the form

$$\left(L_b\right)_{1S}[kHz]=L_{1S}[kHz]+13.9+\frac{21}{5}\delta L_{2S}[\mathrm{kHz}]; \qquad (54)$$

the PBFT correction to 2*S* Lamb shift $\delta L_{2S}$ will be found below in sub-section 4.5, where the value (54) will be calculated.

# 4. Corrections to the atomic energy levels and comparison with experiment

In this section we analyze the hyperfine contributions to the energy levels of light hydrogenlike atoms, where the discrepancy between theory and experiment exceeds the uncertainties in their determination and apply the appropriate corrections of PFBT derived above. We show that the corrections of PBFT provide a perfect conformity between theoretical and experimental values for all parameters listed in the introduction section: 1*S*-2*S* interval in positronium (sub-section 4.1); spin-spin splitting in positronium (sub-section 4.2), proton charge radius derived from the classic 2*S*-2*P* Lamb shift (sub-section 4.3), proton charge radius derived from the ground state Lamb shift in hydrogen (sub-section 4.5). We also pay a separate attention to the 2*S*-2*P* Lamb shift in $He^+$ (sub-section 4.4).

### 4.1. 1*S*-2*S* interval in positronium.

Modern theoretical value of this interval is [1]

$$E_{1S-2S}^{\mathrm{Ps}}=1\ 233\ 607\ 222.2(6)\ \mathrm{MHz}, \qquad (55)$$

and the most precise result of experimental measurements is as follows:

$$1\ 233\ 607\ 216(1) \text{ MHz [2].} \qquad (56)$$

One can see that the deviation between the values (55) and (56) more than three times larger than the uncertainty of measurement of 1*S*-2*S* interval.

Now we introduce the PBFT correction to 1*S*-2*S* transition as the sum of eqs. (34c) and (35):

$$\left(\delta W_b\right)_{total}^{\mathrm{Ps}}(1S-2S) = \delta W_b^{\mathrm{Ps}}(1S-2S) + \left(\delta W_b\right)_{ann} = 6.35 \text{ MHz}.$$

Hence the 1*S*-2*S* interval in positronium corrected in PBFT becomes

$$\left(E_b\right)_{1S-2S}^{\mathrm{Ps}} = E_{1S-2S} - \left(\Delta W_b\right)_{1S-2S}^{\mathrm{Ps}} = 1\ 233\ 607\ 215.8(6) \text{ MHz,} \qquad (57)$$

which perfectly agrees with the experimental value (56).

### 4.2. Spin-spin interval in hydrogenlike atoms

In sub-section 3.2 we have shown that the correction of PBFT to hyperfine spin-spin interaction occurs quite negligible for atoms with $m \ll M$ (*e.g*., hydrogen, eq. (44) and muonium, eq. (45)). For positronium we derived eq. (46), which now will be used for the comparison with experimental data.

The theoretical value of hyperfine splitting in positronium is [1]

$$W_{s-s}^{\mathrm{Ps}} = 203\ 391.7(8) \text{ MHz,} \qquad (58)$$

which does exceed the corresponding experimental data 203 389(2) [20] and 203 387(2) [21].

Eq. (46) allows us to compute the corrected PBFT value of hyperfine spin-spin interval in positronium, using the numerical value (58):

$$\left(W_b\right)_{s-s}^{\mathrm{Ps}} = 203\ 386(1) \text{ MHz.} \qquad (59)$$

This result is already in a good agreement with the experimental data.

### 4.3. $2S_{1/2}$-$2P_{1/2}$ Lamb shift in hydrogen.

It is well known that the dominant problem of exact theoretical evaluation of the classic Lamb shift $L_{2S-2P}$ is the uncertainty arising from the proton charge radius $r_p$. Due to this reason many authors reverse the problem, and estimate $r_p$ from the obtained data on $L_{2S-2P}$ shift (see, *e.g.*, [1]). It is also known that the estimated value of $r_p$ via the measurement of classic Lamb shift systematically exceeds the magnitudes of $r_p$, obtained in the electron-proton scattering data and other methods for evaluation of $r_p$ in physics of elementary particles [4]. This prompted scientists to assume [22] that the uncertainties in estimation of $r_p$ in the experimental particle physics are significantly underestimated. However, the very recent estimation of proton charge radius via the measurement of 2*S*-2*P* Lamb shift in muonic hydrogen gives the value $r_p$=0.84184(67) fm [5], which is substantially lower than the CODATA value $r_p$=0.8768(69) fm [6]. It is also important that for muonic hydrogen the nuclear size effect contributes significantly (about 2 %) to the 2*S*-2*P* Lamb shift and thus this new value of $r_p$ can pretend to be the most precise result amongst all published.

Below we will show that the PBFT correction (50a) to the 2$S$-2$P$ Lamb shift removes the exiting remarkable disagreement between the estimation of $r_p$ from the classic Lamb shift data and estimation for muonic hydrogen.

First we determine the factor $\gamma_{m2}$, which for hydrogen and muonic hydrogen atoms has the value $\gamma_{m2} = \left(1-\alpha^2/4\right)^{-1/2}$=1.0000066. For muonic hydrogen, where the nuclear size effect contributes significantly to the total 2$S$-2$P$ energy interval, the corrected Lamb shift (50) with the factor $\gamma_{m2}$ computed right above does not practically affect the proton charge radius estimated in ref. [5]. In particular, using the parameterization (1) of ref. [5] for the 2$S$-2$P$ energy difference, one can show that the correction (50a) influences the estimated proton size in the order of magnitude $10^{-4}$ fm, which is below of the measurement uncertainty [5]. In contrast, for the hydrogen atom the correction (50a) and finite nuclear size effect have comparable values, and the proton charge radius derived with and without correction (50a) acquires a difference to be substantially larger than the measured/calculated uncertainty.

In order to estimate the proton charge radius from the classic Lamb shift, we use the parameterization

$$L_{2S-2P}\left(r_p\right) = A + B r_p^{\,2}, \qquad (60)$$

which is based on the known fact [3] that the term proportional to $r_p^{\,2}$ is additive. Here $A$ and $B$ are the coefficients, whose numerical values can be found via common calculation of 2$S$-2$P$ Lamb shift in hydrogen [3] for different values of proton charge radius [23, 24]:

$$A\text{=1057695.05 kHz}, \; B\text{=195.750 kHz/fm}. \qquad (61a\text{-}b)$$

In the framework of PBFT, the eq. (60) is appropriately modified:

$$\left(L_{2S-2P}\right)_{PBFT} = \gamma_{m2}^{\;2} A + \gamma_{m2}^{\;2} B \left(r_{PBFT}\right)_p^{\,2}, \qquad (62)$$

where $\left(r_{PBFT}\right)_p$ is the proton size predicted by PBFT. Putting in eq. (60)

$$r_p\text{=0.876(6) fm} \qquad (63)$$

(CODATA value [6]), and equating (60) and (62), we derive the expression for the new proton size evaluated in PBFT:

$$\left(r_{PBFT}\right)_p = \sqrt{\frac{A\left(1-\gamma_{m2}^{\;2}\right)}{B\gamma_{m2}^{\;2}} + \frac{r_p^{\,2}}{\gamma_{m2}^{\;2}}} \approx \sqrt{-\left(Z\alpha\right)^2 \frac{A}{4B} + r_p^{\,2}} \qquad (64)$$

with the sufficient accuracy of calculations. Using the numerical values (61) and (63), we obtain $\left(r_{PBFT}\right)_p = 0.834(6)$ fm.

This estimation is much closer to the proton size derived in ref. [5], than the CODATA value [6]. At the same time, now we recall that the modern CODATA value (63) incorporates the experimental data in both particle physics and atomic physics, and, in general, is less than the proton size derived from the classic Lamb shift solely. In particular, the modern data on 2$S$-2$P$ Lamb shift in hydrogen obtained by various authors within the common approach (see refs. [1, 3] and references therein) define the range of variation of the values of $r_p$ between 0.875 fm and 0.891 fm. Thus taking the midpoint $r_p$ =0.883 fm, we obtain

$$\left(r_{PBFT}\right)_p = 0.841(6)\ \text{fm}, \qquad (65)$$

which exactly coincides with the new proton size[1] (i.e. $r_p$=0.84184(67) fm).

### 4.4. $2S_{1/2}$-$2P_{1/2}$ Lamb shift in He$^+$.

Modern computed value of this shift is equal to [3]

$$L^{\mathrm{He}}_{2S-2P}\text{=14 041.46(3) MHz}, \qquad (66)$$

which after the correction (50) becomes

$$\left(L_b^{He}\right)_{2S-2P} = \left(\gamma_{m2}^{He}\right)^2 L^{He}_{2S-2P}\ \text{=14 042.21(3) MHz} \qquad (67)$$

(where $\gamma^{\mathrm{He}}_{m2}$=1.0000266 for He$^+$). The result of measurement of the Lamb shift by an anisotropy quenching method reported in [25], is

$$\left(L^{\mathrm{He}}_{\exp}\right)_{2S-2P}\text{=14 042.52(16) MHz}, \qquad (68)$$

which disagrees with both estimations (66) and (67).

The obvious discrepancy of the experimental value (68) and QED prediction (66) stimulated further experimental research of the 2*S*-2*P* Lamb shift in He$^+$. In course of their work the authors of ref. [25] redesigned a photon detector system to eliminate a residual polarization sensitivity of the photon detectors, which, in authors' opinion, distorted the result of the previous measurement (68). Having implemented this improvement, they reported in [26] a new result

$$\left(L^{\mathrm{He}}_{\exp}\right)_{2S-2P}\text{=14 041.13(17) MHz}, \qquad (69)$$

which again is in disagreement with the alternative predictions (66) and (67).

Thus, the performance of new high precision experiments on the subject appears to be highly required.

### 4.5. 1*S* Lamb shift in hydrogen.

Having corrected the classic 2*S*-2*P* Lamb shift in PBFT, we are now in the position to complete the PBFT corrections to the ground state Lamb shift (54). To the accuracy sufficient for further calculations, we adopt that the term $\delta L_{2S}$ in eq. (54) is completely determined by the corresponding correction to the coefficient *A* in eq. (61a) for 2*S*-2*P* Lamb shift. Thus, we put $\delta L_{2S} \approx \delta A = \left(\gamma_{mn}{}^2 - 1\right)A$. Hence $\delta L_{2S}$=14.08 kHz. Inserting this value into eq. (52), we obtain

$$\left(L_b\right)_{1S} = L_{1S} + \ 73.0\ \text{kHz}. \qquad (70)$$

Our next goal is to estimate the proton charge radius derived from the ground state Lamb shift corrected by eq. (70) via the comparison of calculated and experimental data on 1*S* Lamb shift collected in Table 12.3 of ref. [3]. One should point out that major part of experiments for the measurement of the 1*S* Lamb shift in hydrogen has been carried out with standard radiofrequency method, whose data are rather widely scattered between the values 8 172 798 kHz and 8 172 874 kHz, with the typical measurement error about 30-50 kHz. In these conditions we select the result of the mentioned Table [3]

$$L_{1S} = 8\ 172\ 837(22)\ \text{kHz}. \qquad (71)$$

obtained with Doppler-free two-photon laser spectroscopy [18], which provides more accurate determination of both 2*S*-2*P* and 1S Lamb shifts in comparison with radiofrequency method.

The same work [3] presents the corresponding theoretical values $(L_{1S})$ for two different values of proton charge radius:

$$L_{1S} = 8\ 172\ 663(6)\ \text{kHz}\ (r_p = 0.805(11)\ \text{fm})\ [23],\ \text{and} \qquad (72a)$$

$$L_{1S} = 8\ 172\ 811(14)\ \text{kHz}\ (\text{for}\ r_p = 0.862(12)\ \text{fm})\ [24]. \qquad (72b)$$

According to eq. (70), we have to add 73.0 kHz to the values (72a-b). Hence we obtain

$$(L_b)_{1S} = 8\ 172\ 736(14)\ \text{kHz}\ (r_p = 0.805(11)\ \text{fm}),\ \text{and} \qquad (73a)$$

$$(L_b)_{1S} = 8\ 172\ 884(14)\ \text{kHz}\ (\text{for}\ r_p = 0.862(12)\ \text{fm}). \qquad (73b)$$

Using the parameterization (60) with the corrected data (73a-b), we find the coefficients $(A_b)_{1S}$ and $(B_b)_{1S}$ as follows: $(A_b)_{1S}$=8 171 727.2 kHz, $(B_b)_{1S}$=1 557.58 kHz/fm. This allows us to determine the proton charge radius, equating to each other the corrected theoretical value and the experimental value (71). This coincidence occurs at

$$r_p = 0.844(22)\ \text{fm}. \qquad (74)$$

This estimation again perfectly agrees with the value of $r_p$ determined through the 2*S*-2*P* Lamb shift in PBFT (65) and with the proton size determined in ref. [5].

# 5. 2*S*-2*P* interval in Li-like uranium

We emphasize that our approach is well applicable not only to light hydrogenic atoms, but also to the entire atomic physics, including the case of heavy multi-charged ions. Thus the corrections of PBFT we have introduced are relevant for both light and heavy atoms. The basic reason, distinguishing the cases of light and heavy atoms from mathematical viewpoint, is that for the latter case the parameter $Z\alpha$ is not small (for example, for uranium it is about 0.67). At the moment PBFT is formulated in the form of perturbation theory based on the Dirac and Breit equations. Translation of PBFT to the QED mathematical language is also possible, but stays outside the scope of the present work. Nonetheless, we can still apply the approach of perturbation theory in the derivation of PBFT corrections for heavy atoms too, at least while the theoretical and experimental uncertainties in estimation of energy intervals in such atoms remain comparably large (from $10^{-3}$ to $10^{-4}$ in relative units). At the same time, the introduction of PBFT corrections to the effects of finite size of nucleus and its polarization requires a separate research. The same remark is relevant with respect to interelectronic interaction for Li-like multi-charged ions, which should be processed within PBFT, too.

Thus a consistent analysis of heavy atoms in the framework of PBFT will be done elsewhere. In the present contribution we can consider, at least qualitatively, the 2*S*-2*P* interval in heavy Li-like uranium measured with the experimental uncertainly about $3\cdot10^{-4}$ (280.59(10) eV [27] and 280.52(10) eV [28]), which remains the most precise result in the physics of multi-charged heavy ions. The current theoretical estimation of this interval is 280.71(10) eV [29] and exceeds the experimental value by about 0.07 %.

Correcting this theoretical value within PBFT, we remind again that our approach does not imply any modifications of QED structure. Rather we introduce the PBFT corrections to the input of resulting QED expressions by the replacements $m \to mb_n$, $U \to \gamma_n U$, and $\boldsymbol{r} = \boldsymbol{r}'/(b_n \gamma_n)$ for the bound electron (see ref. [7]). In addition, we take into account that such corrections even for heavy atoms cannot be large (maximum few percents in relative units for each energy term), and further deal with only the dominant contributions into $2S$-$2P$ interval in heavy Li-like U: interelectronic interaction with the leading term of one-photon exchange ($W_{ee}$=368.83 eV [29]), self-energy and vacuum polarization ($W_{SE}$=-54.98(7) eV [30]), and finite nuclear size correction ($W_{FNS}$=–33,35(6) eV [31]).

*Interelectronic interaction* is described by the Hamiltonian $V_{ee}$ [32], which is added to the Coulomb interaction between the nucleus and electron, so that we obtain

$$H = -\frac{Ze^2}{r} + V_{ee} = -\frac{Ze^2}{r}\left(1 - V_{ee} r / Ze^2\right). \qquad (75)$$

In PBFT the Hamiltonian of interelectronic interaction has the form of (75) in $r'$-coordinates. Hence one can see that the maximum correction to this interaction induced by PBFT has the value of

$$\delta W_{ee} \approx W_{ee}\left(1 - b_n \gamma_n\right)/Z \qquad (76)$$

due to the scaling transformation (9). For $n$=2 state of uranium ($Z$=92), the parameter $\left(1 - b_{mn} b_{Mn} \gamma_{mn} \gamma_{Mn}\right)/Z$=6.3·10$^{-4}$, and the correction to the interelectronic interaction becomes

$$\delta W_{ee} = 368.83 \cdot 6.3 \cdot 10^{-4} = 0.23 \text{ eV}. \qquad (76a)$$

The sign of $\delta W_{ee}$ is positive, because the transformation (9) signifies that at the semi-classical level the increase of the radius of the orbit of $1S$ electrons in Li-like uranium, which, in turn, does increase the screening Coulomb effect on $2S$ and $2P$ states of this atom.

The correction $\delta W_{SE}$ to *self-energy* and *vacuum polarization* within PBFT is determined straightforwardly via eq. (50a) for the effective nuclear charge $Z' = Z - 2 = 90$. Hence

$$\delta W_{SE} = -54.98\left(\left(1 - (Z'\alpha)^2/4\right)^{-0.5} - 1\right) = \text{-3.23 eV}. \qquad (77)$$

The PBFT correction $\delta W_{FNS}$ *to finite nuclear size* emerges due to the fact that in the classical analogy of PBFT, the electron orbits around the nucleus at larger radius than in the standard theory and thus, it is less sensitive to variation of nuclear size than in the standard approach. In order to estimate this effect numerically, we use the approximate expression for nuclear size effect presented in [31], which in the general form reads:

$$\Delta E_{nS_{1/2}} - \Delta E_{nP_{1/2}} = f(n,(Z\alpha))\left(mZe^2 R/\hbar^2\right)^{\sqrt{1-(Z\alpha)^2}}, \qquad (78)$$

where $f(n,(Z\alpha))$ is some function defined semi-numerically, $R$ is the effective nuclear radius, and $r_B = \hbar^2/mZe^2$ is the Bohr radius. In PBFT, eq. (78) should be modified by the replacement $m \to mb_n$. For $n$=2 state, $1/b_n = \left(1 - (Z\alpha)^2/4\right)^{-1}$, and for the numerical value of rhs of eq. (78) -33.35 eV [31], we obtain the PBFT correction to the finite nuclear size effect as follows:

$$\delta W_{FNS} = -33.35\left(\left(1-(Z\alpha)^2/4\right)^{\sqrt{1-(Z\alpha)^2}} - 1\right) = 2.83 \text{ eV}. \tag{79}$$

Thus the total correction to 2*S*-2*P* interval in Li-like U is determined as the sum of eqs. (76a), (77) and (79):

$$\delta W_{total} = \delta W_{ee} + \delta W_{SE} + \delta W_{FNS} = -0.17 \text{ eV}.$$

Hence the corrected within PBFT 2S-2P interval in Li-like uranium becomes

$$(E_b)_{2S-2P} = E_{2S-2P} + \delta W_{total} = 280.54(10) \text{ eV}, \tag{80}$$

that is in a perfect agreement with both experimental values mentioned above 280.59(10) eV [27] and 280.52(10) eV [28].

At the same time, at this stage it would be hasty to interpret this result as another achievement of PBFT, since the correction to interelectronic interaction has been evaluated only qualitatively. Rather this result demonstrates a heuristic potential of PBFT in its application to heavy atoms. Thus, it seems very interesting to extend consistently the pure bound field theory to the case of heavy atoms, especially with the appreciable success of this theory demonstrated above for the light hydrogenlike atoms, including positronium.

# 6. Conclusion

In this paper we verify the Pure Bound Field Theory (PBFT) based on re-postulated Dirac and Breit equations at the scale of hyperfine contributions to the atomic energy levels. On this way we consistently considered the fine structure corrections, as well as corrections to hyperfine spin-spin interaction and corrections to the Lamb shift.

We have demonstrated that the fine structure corrections have the order of magnitude $mc^2(Z\alpha)^6 m/M$ and scales as $n^{-5}$ and $n^{-6}$. Hence they are practically significant only for 1*S* state of hydrogenlike atoms, in particular, in the re-estimation of ground state Lamb shift in the hydrogen atom. For 1*S*-2*S* interval in positronium, there appears an additional component of correction due to the appropriate modification of annihilation term in PBFT, and both corrections completely eliminate the available discrepancy between theoretical value and experimental data for 1*S*-2*S* interval.

The corrections to be brought by PBFT to the spin-spin interval occur negligible for the hydrogenlike atoms with $m \ll M$ (for hydrogen and heavier atoms, eq. (44), it is due to the negligible value of the correcting factor $(Z\alpha)^2 \frac{2mM}{(M+m)^2} W_{s-s}$ in comparison with calculation uncertainly; for muonium, eq. (45), it is due to cancellation by the same PBFT correction in the magnetic moment to mass ratio value derived from Zeeman effect). At the same time, such PBFT correction acquires significant value for the spin-spin interval of 1*S* state of positronium (eq. (46)), where the correction of PBFT removes the discrepancy between theoretical and experimental results.

The PBFT correction to the Lamb shift emerges due to the replacements (5a-c) in corresponding QED equations. Such a correction is directly applicable to the classic 2*S*-2*P* Lamb shift (eq. (50)), where for the hydrogen atom we observe an exact coincidence of corrected by us theoretical value of proton charge radius $r_p$=0.841(6) fm (eq. (65)) with the latest result of measurement via the 2*S*-2*P* Lamb shift in muonic hydrogen $r_p$=0.84184(67) fm [5]. The obtained correction to 2*S*-2*P* Lamb shift contributes to the corresponding correction to 1*S* Lamb shift via eq. (54), along with the correction of PBFT to 1*S*-2*S* interval (34b). Introducing the PBFT corrections to the 1*S* Lamb shift in hydrogen, we obtained the proton charge radius $r_p$=0.844(22) fm (eq. (74)), which practically coincides with the result yield by the classic 2*S*-2*P* Lamb shift.

We have considered 2*S*-2*P* interval in Li-like uranium and shown that PBFT also gives a better coincidence of theoretical and experimental values within the range of their uncertainty.

In Table 1 we summarize the results of QED without and with the corrections we introduced, in comparison with corresponding experimental data. These data completely support our principal idea to modify the Dirac equation for non-radiative EM field of bound electron, which, in its turn, lead to further modifications of equations of the atomic physics. These modifications induce corrections into the effects to be not directly related to each other, but characterized by the same final result: practical elimination of deviations between theory and experiment.

We emphasize that the fine structure correction of PBFT in positronium and corrections to spin-spin interaction result from the appropriate modification of Breit equation in PBFT. In the analysis of radiative corrections to the atomic energy levels, we consider PBFT as a complementary to QED, and no modifications of QED core structure are implied. Nonetheless, the corrections of PBFT to QED results do emerge due to the replacement of electric potential $U$ by $\gamma_{mn}U$ in the resulting QED expressions.

Further, it would be fair to bring up that this work is initiated based on an idea of the third author that the rest mass of any object bound to a given field should be decreased as much as the mass equivalent of the "static binding energy" coming into play (and this, for classical particles, already at rest) [33, 34]. It is worth to note that, via such an approach, he was able to predict the decay rate retardation effect of bound muons [35]. However, a detailed discussion of this idea falls outside the scope of the present paper.

Table 1

Correction of QED results within PBFT in comparison with corresponding experimental (recommended) values for those parameters, where a high measuring precision has been achieved

| Parameter | QED result | Result corrected by us | Experimental value(s) | Ratio "deviation/uncertainty" before correction | Ratio "deviation/uncertainty" after correction |
|---|---|---|---|---|---|
| 1*S*-2*S* interval in positronium | 1 233 607 222.2(6) MHz | 1 233 607 215.8(6) MHz<br>(sub-section 4.1, eq. (57)) | 1 233 607 216(1) MHz [2] | 3.0· | <1 |
| Spin-spin splitting in positronium | 203 391.7(8) MHz | 203 386(1)<br>(sub-section 4.2, eq. (59)) | 203 389(2) [18]<br>203 387(2) [19] | 1.5<br>2.5 | -1.5<br><1 |
| Spin-spin splitting in muonium | 4 463 302.88(55) | remains non-corrected<br>(sub-section 3.2, eq. (45)) | 4 463 302.78(5) [20] | <1 | <1 |
| Proton charge radius (2*S*-2*P* Lamb shift in hydrogen) | 0.876(6) fm | 0.841(6) fm<br>(sub-section 4.3, eq. (65)) | 0.84184(67) fm [5] | 5.7 | <1 |
| Proton charge radius (1*S* Lamb shift in hydrogen) | 0.876(6) fm | 0.844(22) fm<br>(sub-section 4.5, eq. (74)) | 0.84184(67) fm [5] | 5.7 | <1 |
| 2*S*-2*P* interval in Li-like uranium | 280.71(10) eV | 280.54(10) eV<br>(section 5, eq. (80)) | 280.59(10) eV [26]<br>280.52(10) eV [27] | 1.0<br>1.9 | <1<br><1 |